\documentclass[aip,apl,reprint]{revtex4-1}
\usepackage[latin1]{inputenc}
\usepackage{amsmath}
\usepackage{amsfonts}
\usepackage{amssymb}
\usepackage{bm}
\usepackage{natbib}
\usepackage[dvips]{graphicx}

\begin{document}
\title{Strain balanced quantum posts}
\date{\today}
\author{D. Alonso-\'Alvarez.}
\email{diego.alonso@csic.es}
\author{B. Al\'en}
\author{J. M. Ripalda}
\author{J. M. Llorens}
\author{A. G. Taboada}
\author{F. Briones}
\affiliation{IMM-Instituto de Microelectr\'onica de Madrid (CNM-CSIC), Isaac Newton 8, 28760 Tres Cantos, Spain}
\author{M. A.Rold\'an}
\author{J. Hern\'andez-Saz}
\author{D. Hern\'andez-Maldonado}
\author{M. Herrera}
\author{S. I. Molina}
\affiliation{Departamento de Ciencia de los Materiales e Ing. Metal\'urgica y Q. I. Universidad de C\'adiz, Campus Universitario de Puerto Real,  11510 Puerto Real,  C\'adiz, Spain}


\begin{abstract}
Quantum posts are assembled by epitaxial growth of closely spaced quantum dot layers, modulating the composition of a semiconductor alloy, typically InGaAs. In contrast with most self-assembled nanostructures, the height of quantum posts can be controlled with nanometer precision, up to a maximum value limited by the accumulated stress due to the lattice mismatch. Here we present a strain compensation technique based on the controlled incorporation of phosphorous,  which substantially increases the maximum attainable quantum post height. The luminescence from the resulting nanostructures presents giant linear polarization anisotropy.
\end{abstract}
\maketitle

The versatility of epitaxial techniques has enabled new types of nanostructures beyond quantum wells, wires and dots (QDs), with innovative and useful properties. This is the case of epitaxial quantum posts (QPs, also called columnar QDs and quantum rods), which form  by growing  very closely spaced QDs layers.~\cite{He2004, He2007, Li2007a, Ridha2008} The resulting vertically coupled QDs behave as a single nanostructure. According to the calculations of the electronic structure of QPs made by He ~\textit{et al.}, electrons are delocalized along the whole length of the nanostructure whereas holes are confined in the strained-induced region at its bottom.~\cite{He2007} Large dipole moments can be induced in such situation by further separating electrons and holes with a longitudinal electric field. This results on exceedingly tunable exciton radiative lifetimes, from few ns to tenths of miliseconds at low temperatures. Such feature is of interest for different applications like quantum memories or highly non-linear electro-optical devices.~\cite{Krenner2008} The possibility of controlling the height of the nanostructures with nanometer precision also makes QPs of interest in applications where the linear polarization of the absorbed/emitted light is a key issue. That is the case of semiconductor optical amplifiers (SOAs) where a polarization independent optical gain is desirable. QPs could be such material, combining the broadband amplification, high saturation output power, and ultrafast response expected for QDs SOAs with a control of the light polarization.~\cite{Li2007a, Sugawara2004} Kita \textit{et al.} demonstrated that eight alternating  bilayers of InAs/GaAs are enough to produce QPs showing an inversion from transverse electric (TE) mode dominant emission to transverse magnetic (TM) mode dominant emission.~\cite{Kita2002} That means an aspect ratio (height/diameter) of about 0.8, very difficult, if not impossible, to achieve by self-assembling of a single QD layer. In quantum information technology applications, the requirements of the QPs might be even more demanding and the QPs investigated by Krenner \textit{et al.} had a height of 40 nm, with an aspect ratio of 2.~\cite{Krenner2008} Meanwhile, Li \textit{et al.} also reported QPs with 41 nm height but with an aspect ratio of 4, due to the smaller lateral size of their nanostructures.\cite{Li2008}

In this work we report the application of a strain balanced technique to increase the height of QPs up to 120 nm with an aspect ratio of 9.2. The strain balanced method is based on compensating the compressive stress introduced by the nanostructures with a tensile stress in the spacer between the layers. In this method, alternating compressive/tensile regions reduces the accumulated stress, increasing the critical thickness for plastic relaxation. Strain balanced growth has been applied successfully in quantum wells and QDs, especially in the field of photovoltaics. \cite{Ekins-Daukes1999, Alonso-Alvarez2008}
\begin{figure}
	\includegraphics{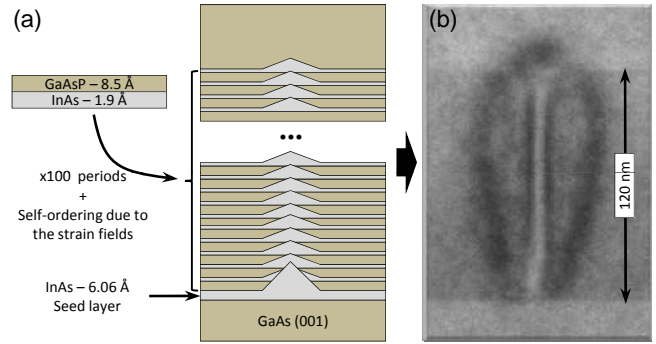} 
	\caption{(a) Schematic representation of the sample and QPs formation. (b) XTEM image of an individual quantum post in the sample under study.}
	\label{Estructura}
\end{figure}

The sample was grown by solid source molecular beam epitaxy (MBE) on GaAs (001). The growth of the QPs begins with a QDs layer formed after the deposition of 2 ML of InAs at 510$\,^{\circ}\text{C}$. On top of it, we grow a short period InAs/GaAsP superlattice at the same substrate temperature. Due to the strain field of the seed QDs layer, In adatoms of the superlattice migrate towards the top of buried QDs whereas Ga adatoms do so in the opposite direction (Fig. \ref{Estructura}(a)). The GaAsP is 8.5 \AA\ thick per period and has a nominal P content of 14\%. We use 1.9 \AA\ of InAs per period and a total number of periods of 100 to complete the QPs growth sequence. Growth rates for the InAs and GaAsP layers are 0.02 ML/s and 0.5 ML/s respectively. The degree of stress compensation in the sample is estimated in 65\% as revealed by \textit{in-situ} accumulated stress measurements (not shown). More specifically, the bending of the substrate due to the accumulated stress was obtained by measuring the deflection of two laser beams reflected on the sample surface.~\cite{Alonso-Alvarez2011b}  A continuous wave Nd:YAG laser frequency doubled to 532 nm was used as the excitation source to investigate the photoluminescence (PL) between 20 K and 300 K. The PL spectrum was recorded with a Peltier cooled InGaAs array attached to a 0.3 m focal length spectrograph.

Fig.~\ref{Estructura}(b) shows a cross sectional transmission electron microscopy (XTEM) image of the sample. Vertical columns are clearly visible and confirm the formation of tall QPs. The average density of dislocations is estimated at 10$^7$ cm$^{-2}$. The QPs height and diameter are $\sim$120 nm and $\sim$13 nm, respectively, giving an aspect ratio of 9.2. The QPs are in effect embedded in a short period InGaAs/GaAsP supperlattice due to the successive alternation of InGaAs wetting layers and GaAsP strain compensating layers, observed in the image with an homogeneous contrast. The contrast between this superlattice and the top and bottom GaAs layers can be clearly seen in the figure. The darker regions around the QPs are mainly related to the strong strain fields present in the structure. 

\begin{figure}
	\includegraphics{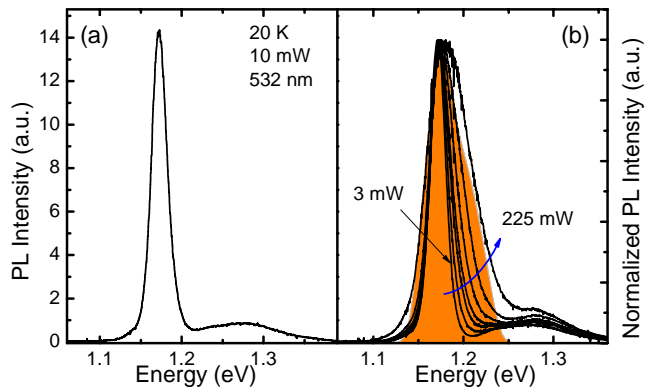} 
	\caption{(a) QPs PL emission spectrum at 20 K and 10 mW. (b) Solid lines: normalized PL spectra as a function of excitation intensity; shaded area: joint DOS calculated as explained in the text (normalized and shifted for greater clarity).}
	\label{PLpower}
\end{figure}

The PL spectrum obtained at 20 K with low excitation power is presented in Fig.~\ref{PLpower}(a). The QPs emission band is centered at 1.17 eV and presents a very low inhomogeneous broadening of 21 meV. The weaker and broader emission band centered at 1.28 eV can be tentatively attributed to small QDs in the seed layer which either do not lead to QPs formation or lead to QPs with a lower In content. The emission spectrum of QPs grown by this method can be shifted to lower energies up to 70 meV by changing the amount of deposited InAs per period from 1.9 \AA\ to 2.2 \AA. Yet, the accumulated elastic energy also produces a larger number of defects reducing the PL intensity by a factor four and increasing the inhomogeneous broadening.

We can get further insight into the properties of our QPs through the dependence of the PL spectrum on the excitation power as shown in Fig.~\ref{PLpower}(b).  As the laser intensity increases, the relatively small linewidth of the emission band allows the observation of an asymmetric broadening towards higher energies.  At the maximum power, the asymmetric full width at half maximum reaches 58 meV. Yet, contrarily to QDs with similar inhomogeneous broadening, no additional bands are observed which could be associated to excited states emission.~\cite{Alen2005, Ripalda2007a}  The asymmetric broadening can be explained if the QPs had a density of states midway between a quantum dot and a quantum wire. To this end, we have calculated the joint density of states (DOS) of a QP of the same average height and diameter. The electronic structure has been calculated using the 8$\times$8 $\bm{k}\cdot\bm{p}$ method together with a function expansion as in Ref.~\onlinecite{marzin1994calculation}. An homogeneously strained infinite cylinder has been considered to include the elastic deformation effects.~\cite{grundmann1995inas} To mimic the inhomogeneous broadening of the spectrum, the resulting joint DOS is artificially broadened with a 20-meV-broad Gaussian convolution and plotted as a shaded area behind the experimental curves in Fig.~\ref{PLpower}(b). To reproduce the experimental broadening at high power we just need to consider the first 12 conduction band states and an equivalent number of valence band states. They correspond to excited states of the vertical confinement potential in the first shell of the in-plane one. Our analysis thus reveals that these QPs have to be described as one-dimensional-like nanostructures, rather than elongated QDs.

\begin{figure}
	\includegraphics{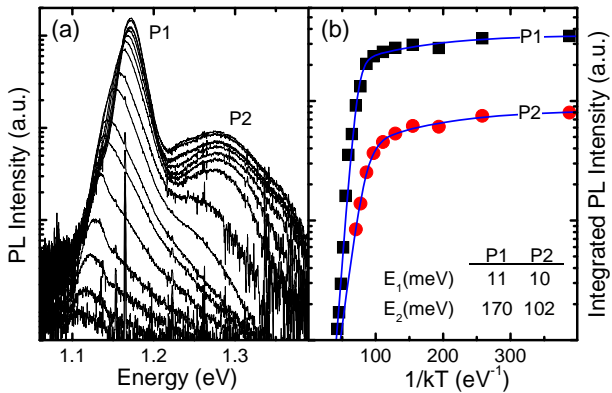} 
	\caption{(a)Temperature dependence of the photoluminescence between 20 and 270 K. The corresponding Arrhenius plot are shown in (b). Continuous thick lines are the best fit to a PL quenching model based on two activation energies. }
	\label{PLvsT}
\end{figure}

The temperature dependence of the PL intensity can be seen in Fig.~\ref{PLvsT}. A Gaussian deconvolution has been applied to study the evolution of the two peaks separately.  The integrated PL intensity is fitted to an Arrhenius type equation with two activation energies.~\cite{Alen2001}  The QPs band (P1) shows an activation energy of 170 meV, whereas for the secondary peak (P2), it is of 102 meV (Fig.~\ref{PLvsT}(b)). In both cases, these energies are compatible with the unipolar escape of carriers from the nanostructures to the matrix.~\cite{Alen2001} The lower activation energy, around 10 meV, must be associated to carrier recombination through defects in the quantum structures or their interfaces.

\begin{figure}
	\includegraphics{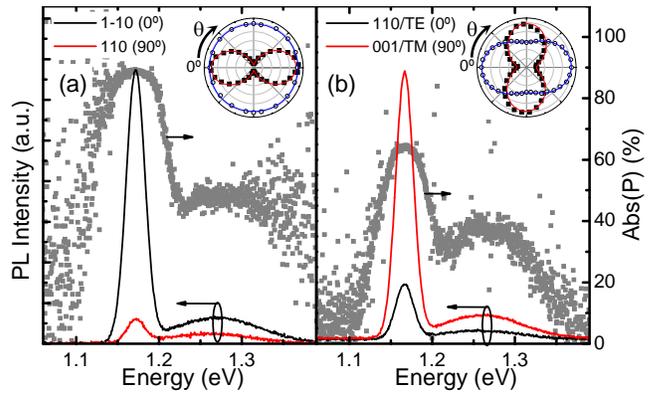} 
	\caption{(a) PL spectrum collected in the growth direction and (b) in the sample edge ([1$\overline 1$0] azimuth). The polarization anisotropy degrees have been calculated as explained in the text (grey squares). The insets show polar diagrams of the PL intensity for varying linear polarization angles in our sample (squares) and in a reference sample (circles). The continuous lines are fits to a sinusoidal dependence.}
	\label{Polarizacion}
\end{figure}

Finally, we address the linear polarization properties of the light emitted by strain balanced QPs. Depending on whether we consider the light emitted in the growth direction or though the cleaved edge, we introduce the following linear polarization anisotropy degrees:
\begin{equation}
P_{\text{[001]}} = \frac{I_{[1\overline 10]}-I_{[110]}}{I_{[1\overline 10]}+I_{[110]}}\quad\text{and}\quad P_{\text{edge}} = \frac{I_{\text{TE}}-I_{\text{TM}}}{I_{\text{TE}}+I_{\text{TM}}}, 
\label{eq:P}
\end{equation}
respectively. In Fig.~\ref{Polarizacion}(a), two PL spectra collected along the vertical axis of the QPs are shown with polarization vectors along the [1$\overline 1$0] and [110] crystal directions, respectively. The inset shows the corresponding polar diagram obtained varying the polarization angle in our sample (solid squares) and in a reference sample containing standard InAs QDs (hollow circles). As it can be observed, the QPs luminescence is strongly linearly polarized along the [1$\overline 1$0] direction, P$_{\text{[001]}}$ = 84\%. The polarization anisotropy degree drops as we deviate from the main emission peak. For the secondary peak, although still noticeable, the anisotropy is half the value at the maximum. Similarly, in Fig.~\ref{Polarizacion}(b) we show the polarization properties of light emitted through one of the sample edges, along the [1$\overline 1$0] direction. As it can be seen, the TM mode is dominant in the whole wavelength range, leading to a maximum P$_{\text{edge}}$=-63\% in the QPs band which drops to -40\% in the secondary band.

The axis of the QPs defines a preferential direction along which there is a greater polarizability of the electron-hole dipole.~\cite{Krenner2008} Therefore, TM polarized emission can be expected for light emitted through the edge, as we observe experimentally along the [1$\overline 1$0] azimuth (P$_{\text{edge}}$=-63\%).~\cite{Kita2002} This is the expected behaviour in QPs, which have a vertical, elongated shape and whose strain properties lead to a hole ground state with large light hole (LH) projection.~\cite{andrzejewski2010} Yet, the prediction fails in the [110] azimuth where P$_{\text{edge}}$ is 27\% (not shown), meaning that the TE component dominates over the TM component for light emitted through the edge in such direction. 

The differences observed between both edges is in agreement with the large polarization anisotropy observed in the growth direction (P$_{\text{[001]}}$ = 84\%). We have identified various factors that can contribute to this result. It is well known that InAs/GaAs QDs usually have an elongated shape in the [1$\overline 1$0] direction,~\cite{Costantini2006} Such effect can contribute to the large  P$_{\text{[001]}}$ value but there is no precedent in the literature for such an effect to produce an 84\% percent degree of polarization by itself (commonly it is between 15\% and 30\% at most).~\cite{Hong-Wen1999}  Another contribution is the self-lateral ordering and composition fluctuations that take place in short period superlattices and some ternary compounds.~\cite{Hong-Wen1999, Pearah1994, Millunchick1997} It is found that the strong anisotropic strain in these structures affects the valence bands mixing in QDs embedded in them, leading to a polarization anisotropy of up to 40\% for light emitted in the growth direction, even with almost symmetric QDs.~\cite{Hong-Wen1999} Finally, an important contribution might be related to the particular microscopic structure of the interfaces in the superlattice.  It has been shown that the bonds present in no common atom interfaces (such as GaP/InAs or InAs/GaSb) lack of a rotoinversion fourfold symmetry axis.~\cite{Krebs1998, Li2010} This property is behind the polarization anisotropy of up to 50\% that is often observed for light emitted perpendicular to such interfaces in superlattices. This effect would scale with the number of interfaces and, hence, it might be of special relevance in our case.

In conclusion, we have substantially increased the maximum attainable height of vertical quantum posts by reducing the accumulated stress through the controlled incorporation of phosphorous in the semiconductor matrix. The optical characterization supports the existence of a 1D density of states in the QPs. Giant linear polarization  anisotropy up to 84\% has been found for light emitted in the growth direction, along the vertical quantum post axis.

We acknowledge the financial support by MICINN (TEC2008-06756-C03-01/03, ENE2009-14481-C02-02, CSD2006-0004, CSD2006-0019, TEC2008-06756-C03-02/TEC and CSD2009-00013), CAM (S2009ESP-1503, S2009/ENE-1477), CSIC (PIF 200950I154) and  J. And. (PAI research group TEP-120 and P08-TEP-03516). 

%
\end{document}